\documentclass[trackchanges,twocolumn,twocolappendix]{aastex701}

\graphicspath{{./}{figures/}}

\begin{document}

\title{A stellar bar hidden in an extreme gas-rich disk galaxy at $z=4.055$}

\author[orcid=0000-0002-3952-8588]{Leindert A. Boogaard}
\affiliation{Leiden Observatory, Leiden University, PO Box 9513, NL-2300 RA, Leiden, The Netherlands}
\email[show]{boogaard@strw.leidenuniv.nl}

\author[orcid=0000-0001-6820-0015]{Luca Costantin}
\affiliation{Centro de Astrobiolog\'{\i}a (CAB/CSIC-INTA), Ctra. de Ajalvir km 4, Torrej\'on de Ardoz, E-28850, Spain}
\email[show]{lcostantin@cab.inta-csic.es}

\author[orcid=0000-0002-1081-883X]{Thor Tepper-Garc\'{i}a}
\affiliation{Sydney Institute for Astronomy, School of Physics A28, The University of Sydney, Sydney, NSW 2006, Australia}
\email[]{thorsten.teppergarcia@sydney.edu.au}

\author[orcid=0000-0001-7516-4016]{Joss Bland-Hawthorn}
\affiliation{Sydney Institute for Astronomy, School of Physics A28, The University of Sydney, Sydney, NSW 2006, Australia}
\email[]{jonathan.bland-hawthorn@sydney.edu.au}

\author[orcid=0000-0001-6586-8845]{Jacqueline A. Hodge}
\affiliation{Leiden Observatory, Leiden University, PO Box 9513, NL-2300 RA, Leiden, The Netherlands}
\email[]{hodge@strw.leidenuniv.nl}

\author[orcid=0000-0002-5247-6639]{Cheng-Lin Liao}
\affiliation{Leiden Observatory, Leiden University, PO Box 9513, NL-2300 RA, Leiden, The Netherlands}
\email[]{liao@strw.leidenuniv.nl}

\author[orcid=0000-0001-9626-9642]{Mahmoud Hamed}
\affiliation{Centro de Astrobiolog\'{\i}a (CAB/CSIC-INTA), Ctra. de Ajalvir km 4, Torrej\'on de Ardoz, E-28850, Spain}
\email[]{mhamed@cab.inta-csic.es}

\author[orcid=0000-0002-9090-4227]{Luis Colina}
\affiliation{Centro de Astrobiolog\'{\i}a (CAB/CSIC-INTA), Ctra. de Ajalvir km 4, Torrej\'on de Ardoz, E-28850, Spain}
\email[]{colina@cab.inta-csic.es}

\author[orcid=0000-0003-4793-7880]{Fabian Walter}
\affiliation{Max-Planck-Institut f\"ur Astronomie, K\"{o}nigstuhl 17, Heidelberg, D-69117, Germany}
\email[]{walter@mpia.de}

\author[orcid=0000-0002-4287-1088]{Oscar Agertz}
\affiliation{Lund Observatory, Division of Astrophysics, Department of Physics, Lund University, Box 118, Lund, SE-221 00, Sweden}
\email[]{oscar.agertz@fysik.lu.se}

\author[orcid=0000-0001-8068-0891]{Arjan Bik}
\affiliation{Department of Astronomy, Oscar Klein Centre, Stockholm University,, AlbaNova University Centre, Stockholm, 106 91, Sweden}
\email[]{arjan.bik@astro.su.se}

\author[orcid=0000-0003-2119-277X]{Alejandro Crespo G\'omez}
\affiliation{Space Telescope Science Institute (STScI), 3700 San martin Drive, Baltimore, MD, 21218, USA}
\email[]{acrespo@stsci.edu}

\author[orcid=0000-0002-3331-9590]{Emanuele Daddi}
\affiliation{CEA, IRFU, DAp, AIM, Universit\'{e} Paris-Saclay, Universit\'{e} Paris Cit\'{e}, Sorbonne Paris Cit\'{e}, CNRS, Gif-sur-Yvette, F-91191, France}
\email[]{edaddi@cea.fr}

\author[orcid=0000-0002-4872-2294]{Georgios E. Magdis}
\affiliation{Cosmic Dawn Center (DAWN), Jagtvej 128, DK2200, Copenhagen, Denmark}
\affiliation{DTU-Space, Technical University of Denmark, Elektrovej 327, DK2800, Kgs. Lyngby, Denmark}
\email[]{geoma@space.dtu.dk}

\author[orcid=0000-0003-4528-5639]{Pablo G. P\'{e}rez-Gonz\'{a}lez}
\affiliation{Centro de Astrobiolog\'{\i}a (CAB/CSIC-INTA), Ctra. de Ajalvir km 4, Torrej\'on de Ardoz, E-28850, Spain}
\email[]{pgperez@cab.inta-csic.es}

\author[orcid=0000-0003-4891-0794]{Hannah \"{U}bler}
\affiliation{Max-Planck-Institut f\"ur extraterrestrische Physik, Gie{\ss}enbachstra{\ss}e 1, Garching, D-85748, Germany}
\email[]{hannah@mpe.mpg.de}

\author[orcid=0000-0003-4678-3939]{Axel Wei\ss}
\affiliation{Max-Planck-Institut f\"ur Radio Astronomie, Auf dem H\"{u}gel 69, Bonn, D-53121, Germany}
\email[]{aweiss@mpifr-bonn.mpg.de}

\begin{abstract}

  The classical picture for the formation of stellar bars---key dynamical drivers of the evolution of galaxies---is through secular evolution of instability in gas poor, stellar-dominated disks. The detection with the James Webb Space Telescope (JWST) of stellar bars and spiral arms in galaxies at early cosmic times has thus challenged $\Lambda$CDM-based expectations, which recent studies reconcile by suggesting that these galaxies are baryon-dominated and have already consumed most of their gas. Yet, a paradox arises, as early galaxies are expected to be increasingly rich in gas, which is generally considered to prevent or slow down stellar bar formation. Here, we show the detection of a stellar bar in GN20, a gas-rich star-forming disk galaxy at a redshift of $z$=4.055, only 1.5 billion years after the Big Bang.  Simultaneous observations of the stars, gas, and dust reveal that GN20 is indeed baryon-dominated (over dark matter; 72$\pm$34\,\%), but the baryonic mass is largely in the form of gas (74$\pm$25\,\%).  This discovery demonstrates that gas-rich disks do support rapid stellar bar formation in the early Universe, motivating a new theoretical perspective on bar formation in gas-rich systems, and providing a potential new mechanism for very early galaxy assembly and quenching.

\end{abstract}

\keywords{\uat{Barred spiral galaxies}{136}; \uat{Galaxy evolution}{594}; \uat{High-redshift galaxies}{734}}

\section{Introduction}

Stellar bars are key dynamical drivers of the evolution of galaxies, including the Milky Way \citep{Johnson1957,deVaucouleurs1964,Blitz1991}. They can efficiently drive flows of gas into the inner regions of a galaxy, where the gas can fuel massive star formation, the build-up of a central bulge, and the growth of a central supermassive black hole \citep{Combes2013, Athanassoula2013, Huang2025}.  The classical picture for the formation of stellar bars---and the general onset of morphological diversity in galaxies---is through the secular evolution of instability in a stellar disk \citep[][]{Sellwood1993}, triggered either internally or by external perturbations.  These secular processes are considered slow and gradual, occurring over many stellar orbits (billion-year timescales in the local Universe), in contrast with the violent and rapid assembly of galaxies at early cosmic epochs \citep[][]{Kormendy2004, Sellwood2014}.

The detection of stellar bars and spiral arms in galaxies in the first two billion years of cosmic time with the James Webb Space Telescope (JWST) \citep[that is, beyond $z\sim3$;][]{Costantin2023, Guo2023, Geron2025, Guo2025, LeConte2025} thus challenged expectations from $\Lambda$CMD-based models \citep{Kraljic2012,Reddish2022}, raising fundamental questions about the mechanisms responsible for the early growth of (disk) galaxies \citep[e.g.,][]{Ferreira2022, HuertasCompany2024, Costantin2025}, and early quenching \citep{Carnall2023, deGraaff2025, Weibel2025}.
In response to these observations, theoretical models showed that stellar bars can form rapidly, in less than a billion-year timescale, if the mass budget of the early disks is dominated by baryons \citep{Bland-Hawthorn2023, Fragkoudi2025}, while new theoretical arguments are emerging for accelerated secular evolution in the early Universe \citep[e.g.,][]{vanderWel2025}.

However, the presence of gas is considered not to be conducive to stellar bar formation, halting, or at least delaying the onset of stellar bars \citep[][]{Bournaud2005, Debattista2006, VillaVargas2010, Athanassoula2013, Bland-Hawthorn2023}.  This has led some recent studies to conclude that the newly-detected high-redshift galaxies with stellar bars must have already consumed their gas \citep[e.g.,][cf. \citealt{Huang2025}]{Costantin2023}, in tension with increasing observational evidence that early galaxies are significantly more gas rich than their local counterparts \citep{Tacconi2010,Tacconi2020, Aravena2024,Decarli2019,Boogaard2023}.

In this paper, we present the discovery of a stellar bar in GN20, a gas-rich, massive disk galaxy at $z=4.055$ \citep{Carilli2010}, only 1.5 billion years after the Big Bang.  Simultaneous observations of the stars, gas and dust show that GN20 hosts a massive ($\approx$\,$5.4\times10^{11}$\,M$_{\odot}$) and dynamically cold disk \citep{Hodge2012, Hodge2015, Colina2023, Ubler2024} that is indeed dominated by baryons---suggesting that the processes capable of assembling and stabilizing disks were already operating efficiently within the first billion years of cosmic history---but, moreover, that the baryons are largely in the form of gas \citep{Boogaard2026}.  This not only constitutes the earliest direct detection of a stellar bar in a measurably gas-rich galaxy to date (cf. \citealt{Amvrosiadis2025, Huang2025}; and without the distorting effects of gravitational lensing), but more importantly demonstrates that stellar bars can in fact develop rapidly in gas-rich, baryon-dominated massive galaxies in the early Universe.  This observational evidence provokes a new perspective on the early formation of stellar bars and structure in galaxies, that is supported by the recent theoretical models from \cite{Bland-Hawthorn2024}.  We use these models to show---now analysing the more massive simulated galaxies, that match more closely to GN20---that stellar bars can form rapidly in massive, gas-rich disks in the early Universe, even when gas dominates the galaxy potential.

\section{Data}\label{sec:data}
\begin{figure*}[ht]
  \centering
  \includegraphics[width=\textwidth]{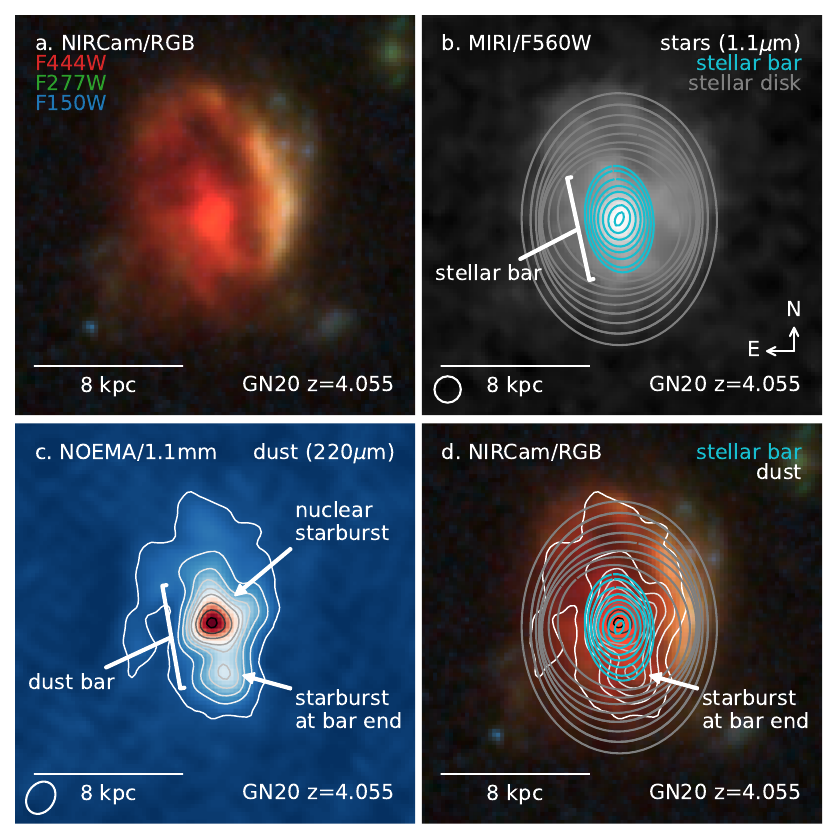}
  \caption{Structure of GN20.  \textbf{(a)} JWST/NIRCam false-color image of the gas-rich
    starburst galaxy GN20 at redshift $z$=4.055. North is up and east
    is left. \textbf{(b)} The 6\,kpc stellar bar in the disk at
    rest-frame 1.1\,$\mu$m light traced by JWST/MIRI, indicated by ellipses of constant brightness. \textbf{(c)} High-resolution sub-mm  observations from the NOrthern Extended Millimeter Array (NOEMA) at rest-frame 220\,$\mu$m reveal the dust extends over the full stellar disk of GN20 \citep{Boogaard2026}, tracing regions of
    strong (obscured) star formation. \textbf{(d)} There is strong
    alignment between the stellar and dust bar.  Clumps of intense
    star formation are visible in the nucleus, fueled by the bar
    funneling material to the galaxy's center, and at the bar-disk
    interface to the south.
    \label{fig:rgb}
  }
\end{figure*}

GN20 features a unique multi-wavelength dataset that covers various tracers of its stellar-, gas-, and dust emission across the spectral energy distribution, from the ultraviolet to the far-infrared and radio.  Here we provide a brief overview of the most recent and relevant observations, referring to \citet[][and references therein]{Boogaard2026} for a more complete overview.

GN20 (R.A.=12$^{\rm h}$37$^{\rm m}$11$^{\rm s}$906, Decl.=+62$^{\circ}$22'12.091; J2000) was observed by JWST \citep{Gardner2023} under program ID~1264 (PI: L.~Colina) as part of the European Consortium MIRI Guaranteed Time and as part of the Galaxy Assembly with NIRSpec Integral Field Spectroscopy (GA-NIFS) survey.  NIRCam imaging observations were taken in the F115W, F150W, F277W and F444W bands \citep{Boogaard2026}, and recalibrated for this work using pipeline v13.0.6 and CRDS 1462 \citep{Bushouse2023}.  MIRI imaging \citep[][]{Rieke2015, Wright2015, Wright2023, Bouchet2015} observations were taken in the F560W, F770W, F1280W, and F1800W bands \citep{Colina2023, Crespo-Gomez2024}. In addition, MIRI Medium Resolution Spectrography \citep[MRS;][]{Wells2015, Argyriou2023} of Paschen\,$\alpha$ is presented in \citet{Bik2024} and NIRSpec Integral Field Spectroscopy \citep[IFS;][]{Jakobsen2022, Boker2022} observations of the H\,$\alpha$ and [N\,\textsc{ii}] lines is presented in \citet{Ubler2024}.  Details about the observational setup and modes are provided in the respective papers \citep{Colina2023, Bik2024, Crespo-Gomez2024, Ubler2024, Boogaard2026}.

GN20 was observed with NOEMA through program W22DU (PI: F. Walter) at 261\,GHz (16\,GHz total bandwidth) in A configuration using all 12 antennas on the longest baselines available (up to 1.7\,km), plus the more compact C configuration to fill in short spacings. Data calibration and imaging were performed in \textsc{gildas}. The final self-calibrated dust continuum map has a spatial resolution (synthesised beam size) of $0.26\times0.2$\,arcsec$^2$, with rms noise of 21.5\,$\mu$Jy\,beam$^{-1}$, and detects the dust continuum at high significance over the entire stellar disk of GN20 (see \citealt{Boogaard2026} for more details).  The total infrared luminosity is $L_{\rm IR}(8-1000\,\mu{\rm m}) \approx 1.6\times10^{13}$\,L$_{\odot}$ \citep[e.g.,][]{Tan2014, Boogaard2026}.  Multi-wavelength cutouts of GN20 are shown in \autoref{fig:rgb}.

\section{Analysis}
\subsection{Morphological Analysis}
\label{sec:morph}
We perform a morphological analysis of GN20 using the MIRI F560W image, which traces the stellar component in the near-infrared regime (rest-frame 1.1~$\mu$m).  The image has a sampling of 0.06 arcsec~px$^{-1}$ and a resolution of 0.2~arcsec (point spread function full width at half maximum; PSF FWHM). The analysis is complemented using NIRCam F444W and MIRI F770W imaging (see \autoref{sec:morph_app}).  At wavelenghts shorter than 4\,$\mu$m, a potential stellar bar feature is not observed, likely due to extreme attenuation \citep[see][]{Boogaard2026}, while at wavelenghts larger than 10~$\mu$m the spatial resolution decreases significantly (\cite{Crespo-Gomez2024}; PSF FWHM $>$ 0.35~arcsec).  Thus, the MIRI F560W image is ideal to detect and characterize a potential stellar bar, tracing the stellar component with minimal contamination, as expected from near-infrared studies in the nearby Universe \citep{Eskridge2000}.  We perform both an isophotal analysis, as well as a Fourier analysis (discussed in \autoref{sec:morph_app}).

\begin{figure*}[ht]
  \centering
  \includegraphics[width=\textwidth]{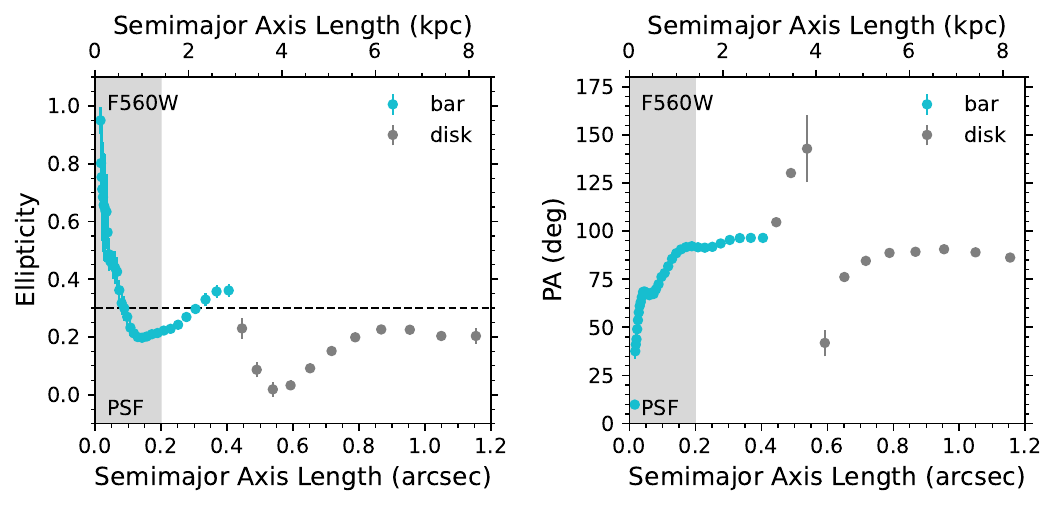}
  \caption{Stellar bar identification. The ellipticity and position angle of the stellar light isophotes at rest-frame $1.1$\,$\mu$m show a clear signature of a stellar bar of approximately $2.8\pm0.1$\,kpc in semi-major axis length, or $5.6\pm0.2$\,kpc full (deprojected) length, that is independently confirmed by Fourier analysis; see \autoref{sec:morph}).  The ellipticity ($\epsilon$) and position angle (PA) of the bar match all four criteria used in literature \citep{Jogee2004, Marinova2007, Costantin2023, Guo2023}: 1.  bar $\epsilon \approx 0.4 > 0.3$ (dotted line), 2.  $\Delta {\rm PA} \approx 5^{\circ} < 15^{\circ}$--20$^{\circ}$ along the bar, 3. $|\epsilon| \approx 0.2 > 0.1$ break at the bar end, 4.  $|\rm PA| \approx 10^{\circ} \geq 10^{\circ}$ between the bar and disk (see \autoref{sec:morph} for more details).
    \label{fig:ellipse_f560w}}
\end{figure*}

For the isophotal analysis, we model the radial surface-brightness profile of GN20 with isophotes (\autoref{fig:ellipse_f560w} and \autoref{fig:ellipse_ext}) using the \textsc{photutils.isophote} astropy package \citep{Jedrzejewski1987, Bradley2020}.  In the first run, we leave all parameters free.  In the second run, we fix the centre of the galaxy to that determined from the innermost isophotes and repeat the fit.  We then characterize the bar and disk in GN20 following the prescriptions detailed in \citet{Jogee2004}, \citet{Marinova2007}, and \citet{Guo2023}.  The results of the fitting can be seen in \autoref{fig:ellipse_f560w}.

In the central (bar-dominated) region, the ellipticity of GN20 rises smoothly to a maximum $\epsilon_{\rm max} \approx 0.4$ and the position angle (PA) remains constant ($|\Delta \rm PA| \approx 5^{\circ}$).  In the outer (disk-dominated) region, the ellipticity drops by $\approx 0.2$ and the PA changes by $\approx10^{\circ}$.  The ellipticity and PA changes surpass the typical threshold for the identification of a stellar bar \citep{Jogee2004, Marinova2007, Guo2023}, though we note the difference in PA between the bar and disk is relatively small.  We estimate the semi-major axis length of the bar, $a_{\rm bar}$, as the radius where the ellipticity reaches its maximum along the bar \citep{Athanassoula2002, Menendez-Delmestre2007, Guo2025, LeConte2025}, finding $a_{\rm bar} = 2.8 \pm 0.1$\,kpc at peak $\epsilon=0.37\pm0.01$, where the corresponding position angle is $96.5^{\circ}\pm1.0^{\circ}$ (measured counter-clockwise from the west).

Given the orientation of GN20, the full deprojected bar length is practically identical to the measured bar length, with a correction factor of $\approx$0.4\,\% following \citet{Yu2022,Kalita2025} using the average disk ellipticity and disk-bar PA difference derived below; smaller than the measurement error.  The mean and standard deviation are computed from the distribution of measured bar lengths from 100 perturbations of the image with the background noise.  It is worth noting, that the above definition could slightly underestimate the true bar length \citep{Martinez-Valpuesta2006, Guo2025}, given the bar typically appears to extend beyond the region of maximal ellipticity, though in GN20 this position is generally close to the radius where the bar ellipticity drops steeply.  We therefore also derive estimates of the bar length as the largest radius at which $\epsilon>0.3$ ($\epsilon>0.2$), which indeed yields a slightly larger bar length of $3.1 \pm0.1$\,kpc ($3.2 \pm0.1$\,kpc).  We also measure the radius of minimum ellipticity after the peak ellipticity \citep{Erwin2005,Athanassoula2002} which provides an upper limit on the bar length, yielding $4.1 \pm0.1$\,kpc.  The median ellipticity and position angle in the region of the bar (taken as the region starting beyond the PSF core, up to the maximum bar length) and disk (taken loosely at the range from 5--8\,kpc; the exact definition does not impact the result), are found to be $\langle{\rm PA}_{\rm bar}\rangle = 97.0\pm1.0$, $\langle\epsilon_{\rm bar}\rangle = 0.36\pm 0.01$, $\langle {\rm PA}_{\rm disk}\rangle = 90.1\pm0.5$, and $\langle\epsilon_{\rm disk}\rangle = 0.20\pm 0.01$.

We repeat the same measurement on the lower-resolution MIRI F770W image at longer wavelengths, as well as the higher-resolution NIRCam F444W image, though we note the analysis of the NIRCam data is more challenging and uncertain due to the complex observed morphology that is possibly still affected by dust attenuation.  We also preform a Fourier analysis on the MIRI F560W image.  Both the multi-wavelength and Fourier analysis yield a consistent bar signature and are discussed in more detail in \autoref{sec:morph_app}.  We adopt the bar length and properties from the isophotal analysis of the MIRI F560W image.

\subsection{Baryon- and gas fraction}
\label{sec:baryon-gas-fraction}

The mass budget of GN20 is well constrained for a galaxy at $z>4$ and measurements of the total dynamical mass (baryons + dark matter) as well as the mass in the baryonic components (stars, gas) are discussed in detail in \autoref{sec:mass}.  Based on these estimates, we infer that GN20 has a significant baryon- or disc fraction (following the definition of \citealt{Bland-Hawthorn2024} and linearly propagating the errors) of $f_{\rm baryon} = f_{\rm disc} = (M_{\rm star} + M_{\rm mol})/M_{\rm dyn} = 0.72 \pm 0.34$, assuming the mass of the gas is dominated by the molecular gas.\footnote{We note that while the measured baryon fraction can exceed unity within uncertainties, as it involves a ratio of three independently measured quantities, the true baryon fraction of the galaxy cannot.}  This implies a gas fraction of $f_{\rm gas} = M_{\rm mol}/(M_{\rm star} + M_{\rm mol}) = 0.74 \pm 0.25$, where we have included an additional 15\% uncertainty in the gas fraction (given the uncertainties discussed in \autoref{sec:mass}).  Including an additional gas component in the form of neutral atomic hydrogen (H\textsc{i}) would further increase the inferred baryon- and gas fraction.  We emphasise that while there are well-known systematic uncertainties involved in the mass estimates of each of the different components of the galaxy, the precise adopted values do not significantly affect the conclusions, consistently yielding a baryon-dominated and gas-rich system.

\section{Results}
\label{sec:results}
The stellar bar in GN20 is discovered in high-resolution JWST imaging with the Mid-Infrared Instrument (MIRI) and Near-Infrared Camera (NIRCam) shown in \autoref{fig:rgb}.  The JWST observations trace starlight at rest-frame near-infrared wavelengths, penetrating the large volume of dust that completely obscures GN20 at shorter wavelengths \citep{Carilli2010, Hodge2012, Hodge2015, Boogaard2026}.  The analysis of the galaxy's surface brightness at rest-frame 1.1~$\mu$m (MIRI 5.6~$\mu$m band) shown in \autoref{fig:ellipse_f560w} provides quantitative evidence of a stellar bar of approximately 6\,kpc in full length (measured along the major axis; \autoref{sec:morph}).  The bar signature is also detected at shorter and longer wavelengths (\autoref{fig:ellipse_ext} in \autoref{sec:morph_app}).  The bar is independently confirmed via a Fourier analysis of the stellar light (\autoref{fig:fourier} in \autoref{sec:morph_app}), which shows a prominent $m=2$ bar mode with a constant phase angle $\phi$ along the bar, and a consistent bar length. The stellar bar was initially not noticed in the MIRI/F560W imaging, but upon closer inspection signatures of the bar can already be seen as a typical bump in the residuals \citep[see][their Fig.~1]{Colina2023} from modelling the light profile with a combination of a central mass concentration and an extended disk component \citep[cf. for example][]{Cuomo2019}.

The stellar bar aligns with a bar-like structure seen in the dust continuum \citep{Hodge2015,Boogaard2026}.  The far-infrared dust emission in GN20 was recently mapped in unprecedented detail with the NOrthern Extended Millimeter Array (NOEMA), which revealed that the dust extends over the full stellar and gas disk \citep[][]{Boogaard2026}.  The far-infrared emission implies a huge star formation rate (SFR) in GN20, exceeding 1000\,M$_{\odot}$\,yr$^{-1}$ \citep{Tan2014,Crespo-Gomez2024}.  Part of this high SFR is likely being driven by the bar funnelling gas and dust into the center, where it triggers an intense nuclear starburst in the gas-rich disk, and fuels the potential active galactic nucleus \citep{Riechers2014,Ubler2024} in the massive gas-rich galaxy.  The enhanced dust emission observed south of the nucleus aligns with the tip of the stellar bar (\autoref{fig:rgb}).  This likely corresponds to enhanced star formation at the bar-disk interface, as seen in nearby galaxies \citep[e.g.,][]{Beuther2017}.  At the same time, the morphology of GN20 also shows some striking differences with the typical morphology of barred galaxies in the nearby universe, including a global asymmetry and one-sidedness of the arm and enhanced star-formation at the bar tip.  In \autoref{sec:discussion} we will show that similar asymmetric features also arise in the simulations.

As a complementary piece of evidence for the presence of a bar, we note there are non-circular motions observed in the ionized gas kinematics, which could be consistent with bar-induced motions \citep[see][incl.\ their Appendix A, for a thorough analysis and detailed discussion of the ionized gas kinematics and possible radial gas flows]{Ubler2024}.  %
We do caution against the interpretation of ionized gas kinematics to confirm the stellar bar, as molecular gas kinematics are be better suited to trace bar structure and the bar-induced inflow of gas (and corresponding mass inflow rate) to the central regions \citep[e.g.,][]{Pastras2025}.  The kinematics of the low-$J$ CO(2--1) emission was studied in great detail already in \citet{Hodge2012} based on very deep observations from the JVLA.  These revealed the velocity field showing a rotating disk in GN20, similar to that now observed in the ionized gas.  However, given the limited CO(2--1) brightness, the deep observations are of insufficient signal-to-noise to detect bar signatures \citep{Hodge2012}.  Future observations of the mid-$J$ CO kinematics may confirm bar-induced motions in the molecular gas.

\section{Discussion \& Conclusions}
\label{sec:discussion}
\subsection{Rapid bar formation in early gas-rich galaxies}
\begin{figure*}[ht]
  \centering
  \includegraphics[width=\textwidth]{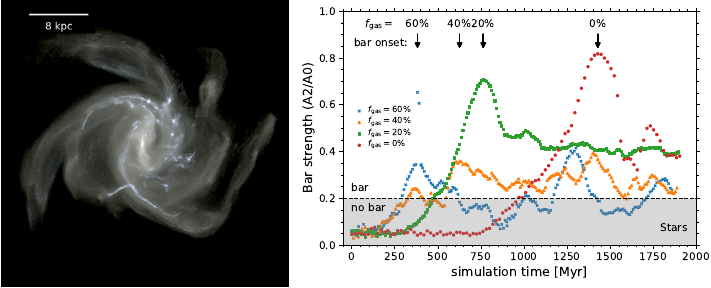}
  \caption{Predicted timescales of stellar bar formation. Theoretical simulations of bar formation in the context of
    large gas fractions from \citep{Bland-Hawthorn2024,
      Bland-Hawthorn2025, Tepper-Garcia2024} produce galaxies with a
    striking resemblance to GN20, featuring a bars and asymmetric
    spiral arm structure, driven by baryon sloshing \citep[][snapshot
    at $f_{\rm gas} = 40$\% shown]{Bland-Hawthorn2025}.  For
    baryon-dominated disks, the simulations predict increased gas
    fractions drive earlier onset of bars, with bar formation in
    massive disks occurring within a few hundred Myrs
    ($f_{\rm baryon} = 0.7$, $M_{\rm halo} = 10^{11}$\,M$_{\odot}$; see \autoref{sec:theory} for more details).
    \label{fig:theory}}
\end{figure*}

The striking discovery of a stellar bar in a gas-rich and baryon-dominated disk galaxy only 1.5 Gyr after the Big Bang motivates an alternative perspective on the rapid assembly of early disk galaxies, in which the gas---and not the stars---takes an increasingly dominant role in shaping the galactic potential \citep{Tepper-Garcia2024, Bland-Hawthorn2024, Bland-Hawthorn2025}.  Critically, classical stellar orbit theory \citep{BinneyTremaine2008} holds only in the low gas-fraction limit.  In contrast, some of the latest theoretical models suggest that stellar bars could indeed form in gas-rich, turbulent disks \citep{Bland-Hawthorn2024}.  These simulations (presented in more detail in \autoref{sec:theory}) show morphologies that are strikingly similar to GN20 (\autoref{fig:theory}), including global asymmetry and one-sidedness potentially due to sloshing of the galaxy in the overall potential.  The sloshing of baryons in such gas-rich systems also offers an alternative scenario for the onset of bar formation \citep{Bland-Hawthorn2025} compared to classical theories where stars dominate the baryon budget and instabilities are driven via interaction \citep{Kormendy2004, Sellwood2014}.  As GN20 features close companions (\citealt{Ubler2024}), both mechanisms could potentially be at play (see \autoref{sec:timescale} for a more extended discussion).

With regard to bar properties, we note that \cite{Bland-Hawthorn2024} found from their intermediate mass simulations (which reach a lower baryon fraction, or disk fraction, than in GN20; $f_{\rm baryon}$=0.5) that, at fixed baryon fraction, the bar length and mass are inversely correlated, which appears to be at odds with the long bar length measured in GN20 (see also tension in \citealt{Huang2025} and \citealt{Amvrosiadis2025}).  Thus, we now analyse the more massive galaxy simulations, which reaches baryon fractions of $f_{\rm baryon}=0.7$ (similar to GN20, cf. \citealt{Huang2025, Amvrosiadis2025}), and find it indeed presents a large stellar bar as seen in GN20 (and also of comparable length; see \autoref{fig:theory_ext}).  Moreover, a key result---shown in \autoref{fig:theory}---is that for a fixed baryon fraction, higher gas fractions appear to drive an earlier onset of the stellar bar, matching the (rapid) timescales required for GN20 (as well as the estimated ages of the stars in the bar of $\approx300$\,Myr; see \autoref{sec:timescale}).  We discuss the underlying theoretical background in more detail in \autoref{sec:theory}.  Overall, this presents striking observational and theoretical evidence that baryon-dominated and gas-rich galaxies can in fact form bars, and moreover that increasingly high gas fractions (expected in the early Universe) can lead bars to form rapidly, on a timescale of only hundreds of Myr (\autoref{fig:theory}), orders of magnitude faster than inferred in the local Universe.

This shift in paradigm, in which at early times not stars but gas dominates the baryon budget and thus drives the early growth of galaxies and their structure, is particularly important, as the last decade of submillimeter observations has shown that the cosmic density of molecular gas was significantly higher in the early Universe \citep{Decarli2016,Decarli2019,Riechers2019,Boogaard2023} and galaxies were significantly more gas-rich \citep[e.g.,][]{Tacconi2010, Tacconi2020, Aravena2024}. Moreover, dynamically cold, molecular gas-rich disks are being detected in galaxies at increasingly early times \citep{Neeleman2020, Rizzo2020, Lelli2021}, even less than a billion years after the Big Bang \citep{Rowland2024}.  It is indeed very likely that most galaxy disks go through a gas-rich phase in their evolution.  We therefore anticipate that, if stellar bars can indeed efficiently form in dynamically-cold, gas-dominated disks, stellar bars may be discovered out to even higher redshifts.  We note a possible stellar bar-like feature was identified in a strongly gravitationally lensed galaxy at $z=4.3$ \citep{Smail2023}.  Also, while, critically, a bar in the gas or dust does not imply a stellar bar, we do note that since the discovery of a dust bar in GN20 a decade ago \citep{Hodge2015}, more bars in dust and gas have been discovered in other massive distant galaxies, also at earlier cosmic times \citep[e.g.,][]{Hodge2019, Gullberg2019, Tsukui2024, Amvrosiadis2025, Huang2025, Umehata2025, Rowland2024}.

\subsection{High-redshift bars driving fast galaxy quenching?}
From a broader perspective, the new framework supports a picture in which the high gas fraction in GN20 could lead the bar to dissolve rapidly (in $<$1~Gyr) into a central bulge \citep[that is, as early as $z\sim3$,][]{Bland-Hawthorn2024}.  Alternatively, if bulge growth is limited (e.g., due to AGN feedback), the stellar bar might also persist on longer timescales, consistent with massive barred galaxies observed at later epochs (some of which also appear to be gas-rich and/or quenched).

If, as the models show, stellar bars can indeed form on short timescales and thereby fuel central star formation effectively in gas-rich galaxies at very early times, this could be a pathway to the formation of the high-redshift (massive) quiescent galaxies that have been discovered with JWST at very high-redshift (\citealt{Carnall2023,deGraaff2025}; even less than a Gyr after the Big Bang, \citealt{Weibel2025}).  In the same way, massive dusty star-forming galaxies like GN20 (known to reside in an overdensity; \citealt{Pope2005, Pope2006, Daddi2009, Carilli2011}), are thought to be the progenitors of massive quiescent galaxies in the local universe, based on their luminous infrared properties and estimated halo masses matching the evolution of high-redshift quasars and local massive spheroidals \citep{Sanders1988, Dudzeviciute2020, Stach2021, Smail2023}.  One of the key prerequisites for the quenching of star formation in the progenitors of massive quiescent galaxies is thought to be compaction and the formation of dense cores or bulges \citep{Zolotov2015, Tacchella2016, Barro2017, Costantin2021, Smail2023}.  Indeed, the stellar bar in GN20 appears to efficiently dissipate gas into the center, driving massive star formation and the build-up of a dense core (\autoref{fig:rgb}).  At the same time, this material could fuel the growth of the central massive black hole, as witnessed by signatures of active black hole activity \citep{Riechers2014, Ubler2024}.  In this respect, GN20 may not just capture a crucial phase in the evolution of massive dusty starburst galaxies at high-redshift towards becoming the massive quiescent galaxies found in the local Universe today, but also highlight how potential bars forming at even higher redshift may be a key mechanism for the rapid assembly and quenching of galaxies at very early cosmic times.

\begin{acknowledgments}
  We thank the referee for a constructive report that helped improve the quality of the paper.
  L.A.B. acknowledges support from the Dutch Research Council (NWO) under grant VI.Veni.242.055 (\url{https://doi.org/10.61686/LAJVP77714}).
L.A.B., J.H. and C.-L.L. acknowledge support from the ERC Consolidator grant 101088676 ("VOYAJ").
The project that gave rise to these results received the support of a fellowship from the “la Caixa” Foundation (ID 100010434). The fellowship code is LCF/BQ/PR24/12050015.
L. Costantin and P.P.G. acknowledges support from grant PID2022-139567NB-I00 funded by the Spanish Ministry of Science and Innovation/State Agency of Research MCIN/AEI/10.13039/501100011033 and by “ERDF A way of making Europe”.
L. Colina acknowledges support by grant PIB2021-127718NB-I00 from the Spanish Ministry of Science and Innovation/State Agency of Research MCIN/AEI/10.13039/501100011033 and by “ERDF A way of making Europe”.
G.E.M. acknowledges the Villum Fonden research grant 13160 “Gas to stars, stars to dust: tracing star formation across cosmic time,” grant 37440, “The Hidden Cosmos,” and the Cosmic Dawn Center of Excellence funded by the Danish National Research Foundation under the grant No. DNRF140.
H.\"{U}. acknowledges support by the Max Planck Society through the Lise Meitner Excellence Program. H.\"{U}. acknowledges funding by the European Union (ERC APEX, 101164796). Views and opinions expressed are however those of the authors only and do not necessarily reflect those of the European Union or the European Research Council Executive Agency. Neither the European Union nor the granting authority can be held responsible for them.
A.B. acknowledges support from the Swedish National Space Administration (SNSA).
O.A. acknowledges support from the Knut and Alice Wallenberg Foundation, The Swedish Research Council (grant 2025-04892), the Swedish National Space Agency (SNSA Dnr2023-00164), the LMK foundation, and eSSENCE-- a Swedish strategic research programme in e-Science.

The JWST NIRCam and MIRI data presented in this article are publicly available from the Mikulski Archive for Space Telescopes (MAST) at the Space Telescope Science Institute. The specific observations analyzed can be accessed via \url{https://doi.org/10.17909/mrb8-x762}.  The IRAM NOEMA observations can be requested via the IRAM Science Data Archive (\url{https://iram-institute.org/science-portal/data-archive/}).

\end{acknowledgments}

\appendix

\section{Additional morphological analysis}\label{sec:morph_app}
\begin{figure*}[ht]
  \centering
  \includegraphics[width=\textwidth]{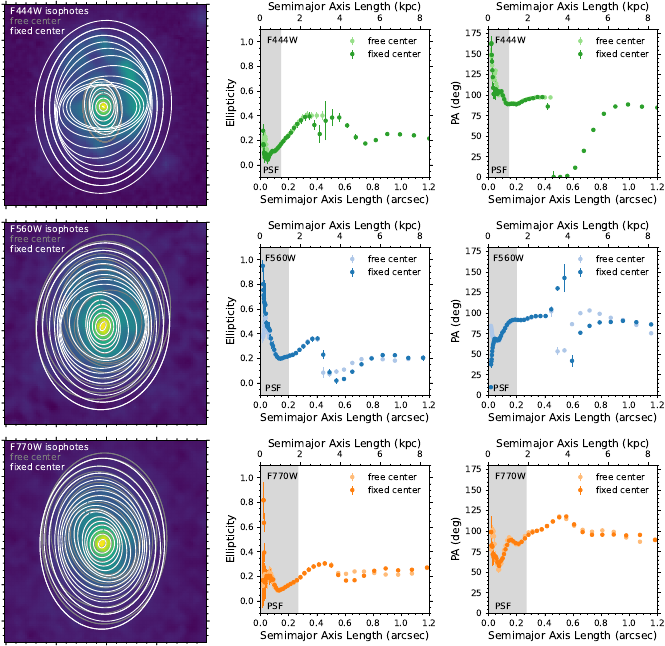}
  \caption{Stellar bar identification at multiple wavelenghts. Isophotal analysis of the stellar bar (same as in \autoref{fig:ellipse_f560w}, with the leftmost panels showing the isophotes), now for the NIRCam F444W (\textbf{top panels}), MIRI F560W (\textbf{middle panels}) and MIRI F770W filters (\textbf{bottom panels}), including an additional fit leaving the center of the isophotes free, yielding consistent results.
    \label{fig:ellipse_ext}}
\end{figure*}

\begin{figure*}[ht]
  \centering
  \includegraphics[width=\textwidth]{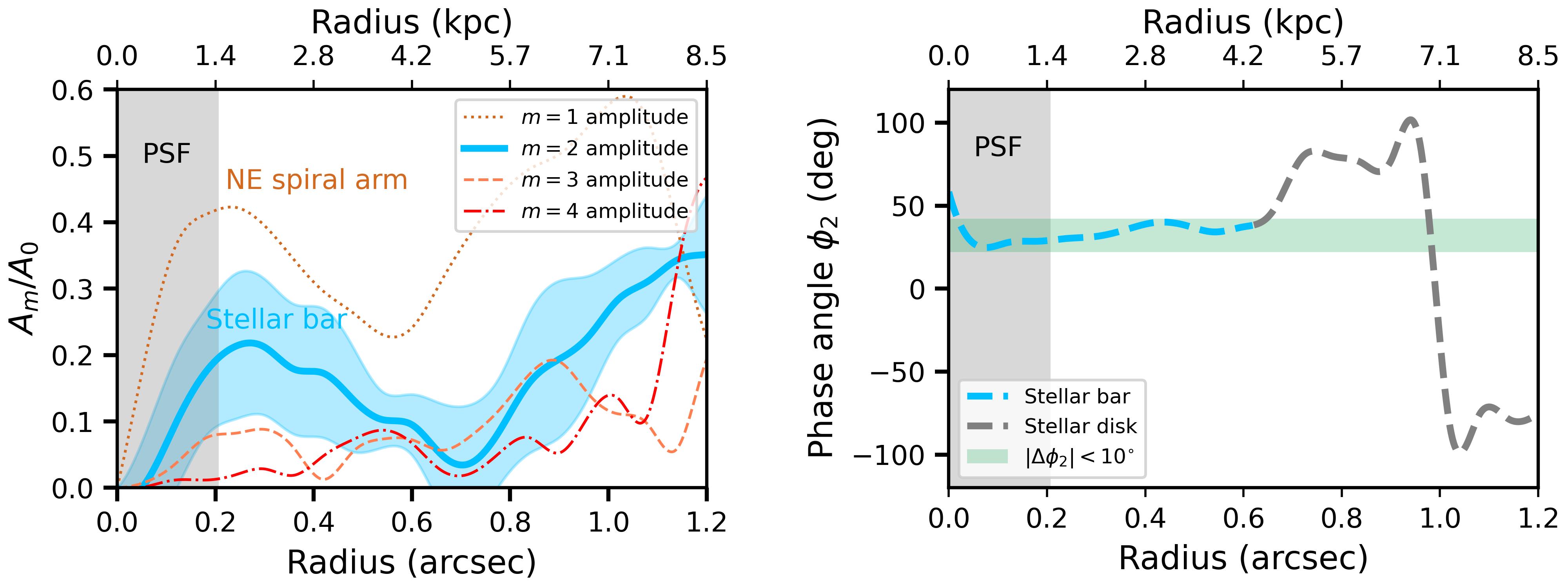}
  \caption{Stellar bar identification in Fourier space.  Two dominant modes are identified ($A_m / A_0 >0.2$),
    which correspond to the NE spiral arm ($m=1$) and the stellar bar ($m=2$).
    \label{fig:fourier}}
\end{figure*}

Here we present additional morphological analysis of the longer- and shorter-wavelength MIRI F770W and NIRCam F444W images, as well as a Fourier analysis on the MIRI F560W data.

We repeat the morphological analysis described in \autoref{sec:morph} on the lower-resolution MIRI F770W filter.  The results can be seen in \autoref{fig:ellipse_ext}.  We find a consistent bar signature with the following bar properties at peak ellipticity: $a_{\rm bar} = 3.2 \pm 0.2$\,kpc, where $\epsilon=0.30\pm0.01$ and the position angle is $110.3^{\circ}\pm3.1^{\circ}$.  For the higher-resolution NIRCam F444W image at shorter wavelengths, we also find a consistent bar signature, with a slightly shorter bar length of $a_{\rm bar} = 2.4 \pm 0.1$\,kpc ($\epsilon=0.40\pm0.01$, $\mathrm{P.A.}=97.3^{\circ}\pm1.3^{\circ}$).  But, we note the isophotal analysis in F444W is more challenging and uncertain due to the complex observed morphology, possibly affected by increasing dust attenuation, and yields orthogonal elliptical isophotes at the bar-disk interface (a second maximum in ellipticity at orthogonal position angle), likely driven by the observed gap in the disk north of the bar and the connection to the arm feature.  In general, the strength of the breaks in ellipticity and position angle between the bar and the disk become smaller at longer wavelengths which is likely the result of the larger PSF at longer wavelengths \citep[e.g.,][]{Liang2024}.

As the nucleus of GN20 is known to be offset from the isophotes \citep{Colina2023}, we also repeat the measurements in both filters leaving the center of the isophotes free, yielding very consistent results (see \autoref{fig:ellipse_ext}; the largest differences are seen in the complex region between the bar and the disk).

For the Fourier analysis, we deprojected the MIRI F560W image and decompose the azimuthal luminosity surface-density distribution into Fourier $m$-components \citep{Ohta1990, Aguerri2000, Buttitta2022}.  The image is stretched along the disk minor axis by a factor of cos($i_{\rm disk}$)$^{-1}$, where $i_{\rm disk} = 40^{\circ}$ represents the inclination of the thick disk derived from the ellipticity profile in the outer region \citep{Padilla2008}.  This inclination is in very good agreement with the value determined by \citep{Ubler2024}.  The radial profile of the $m = 2$ component shows the characteristic behaviour of bars \citep{RosasGuevara2022}, increases with radius and then decreases in the disk region (\autoref{fig:fourier}).  Complementary evidence comes from a constant phase angle of the $m = 2$ component $|\Delta \phi_2| < 10$~deg. From the Fourier analysis, we derived three different estimations of the bar radius: (1) $R_{\rm bar/A_2} = 3.72_{-0.15}^{+0.74}$~kpc from the FWHM of $A_2$, i.e. the outer radius at which $A_2$ decreases to 50\% of its maximum value \citep{Aguerri2000, Athanassoula2002}, (2) $R_{\rm bar/interbar} = 4.17_{-0.60}^{+0.26}$~kpc, from the bar/interbar intensity contrast \citep{Aguerri2000}, and (3) $R_{\rm bar/\phi_2} = 4.38_{-1.48}^{+0.16}$~kpc, from constant $\phi_2$ in the bar region \citep{Debattista2002}. It is worth noticing that the bar length measured through ellipse fitting is shorter than estimations through Fourier decomposion, as
discussed in many studies \citep[e.g.,][]{Athanassoula2002, Erwin2005}. We also derived the strength of the bar $S_{\rm bar} = 0.22^{+0.01}_{-0.02}$ as the maximum of $A_{2, \rm max}$ \citep{Athanassoula2002}.

\section{Baryonic and dynamical mass estimates}\label{sec:mass}

 \begin{figure*}[ht]
  \centering
  \includegraphics[width=\textwidth]{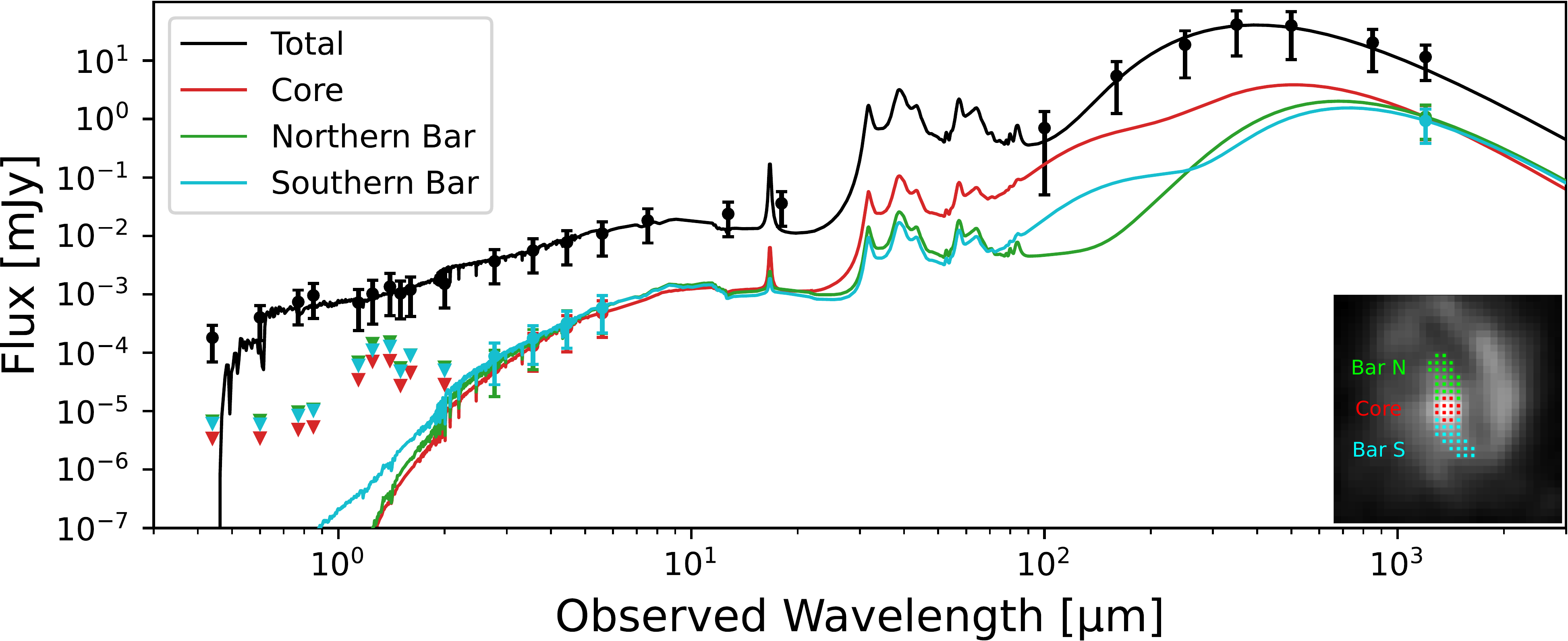}
  \caption{Spectral energy distribution model of GN20.  The best-fit SED model from \texttt{MAGPHYS} with the total, core, northern bar, and southern bar are shown in the black, red, green, and cyan curve, respectively.  The inset shows the regions on a $2''\times2''$ NIRCam/F444W cutout.
    \label{fig:magphys_best-fit}}
\end{figure*}

\begin{table*}
    \centering
    \begin{tabular}{cccccc}
    \hline\hline
         & Total & Total (incl. AGN) & Core & Southern Bar & Northern Bar \\
    \hline
    \multicolumn{6}{c}{total aperture (\textsc{magphys})} \\
        log($M_{\star}/M_{\odot}$) &
        $11.09^{+0.26}_{-0.24}$  &
        $10.98^{+0.27}_{-0.26}$  &
        $10.62^{+0.40}_{-0.53}$ & $10.50^{+0.45}_{-0.51}$ & $10.59^{+0.44}_{-0.55}$ \\
        Mass-weighted age (Myr) & $98^{+159}_{-52}$ &
        $170^{+238}_{-105}$ &
        $295^{+230}_{-183}$ &         $288^{+224}_{-181}$ &  $295^{+218}_{-188}$ \\
        log(SFR/$M_{\star}$\,yr$^{-1}$) & $2.95^{+0.15}_{-0.24}$ &
        $2.67^{+0.26}_{-0.35}$ & $2.09^{+0.38}_{-0.52}$ & $2.00^{+0.39}_{-0.56}$ & $2.08^{+0.39}_{-0.57}$ \\
        $A_{V}$ (mag) &
        $2.21^{+0.42}_{-0.48}$ &
        $1.74^{+0.58}_{-0.60}$ & $5.14^{+1.53}_{-1.62}$ & $4.41^{+1.40}_{-1.45}$ & $4.84^{+1.38}_{-1.52}$ \\
    \hline
    \multicolumn{6}{c}{aperture (\textsc{cigale})} \\

        log($M_{\star}/M_{\odot}$) & $11.06 \pm 0.02$ & $11.06 \pm 0.02$ & $10.56 \pm 0.04$ & $10.44 \pm 0.03$ & $10.49 \pm 0.06$ \\
        Mass-weighted age (Myr) &$60 \pm 2$ & $61 \pm 11$ & $299 \pm 67$ & $257 \pm 32$ & $220 \pm 40$ \\
        log(SFR/$M_{\star}$\,yr$^{-1}$) &  $3.19 \pm 0.02$ & $3.18 \pm 0.02$ & $2.02 \pm 0.04$ & $1.95 \pm 0.03$ & $2.07 \pm 0.06$ \\
        $A_{V}$ (mag) &$2.71 \pm 0.01$ & $2.71 \pm 0.03$ & $4.59 \pm 0.03$ & $4.57 \pm 0.03$ & $4.58 \pm 0.12$ \\
\hline
\multicolumn{6}{c}{pixel-by-pixel (\textsc{magphys})} \\
        log($M_{\star}/M_{\odot}$) & $11.80\pm0.02$ & $\cdots$ & $10.86\pm0.07$ & $10.74\pm0.06$ & $10.91\pm0.06$ \\
        log(SFR/$M_{\star}$\,yr$^{-1}$) & $3.29\pm0.02$ & $\cdots$ & $2.32\pm0.08$ & $2.23\pm0.06$ & $2.37\pm0.06$ \\
    \hline\hline
    \end{tabular}
    \caption{Physical properties derived from aperture photometry with \textsc{magphys} (top section) and \textsc{cigale} (middle section), as well as pixel-by-pixel based SED modelling (\textsc{magphys}; bottom section).  The integrated galaxy properties are obtained using all available multi-wavelength data.  The bulge and bar properties, as well as the pixel-by-pixel-based results are derived using only the PSF-matched HST and JWST filters with wavelengths shorter than MIRI/F560W and the NOEMA 1.1\,mm continuum.
      \label{tab:sed_aperture}}
\end{table*}

The total (baryonic + dark matter) dynamical mass estimate of GN20 is
$M_{\rm dyn} = 5.4 \pm 2.4 \times 10^{11}$\,M$_{\odot}$  based on $^{12}$CO, H\,$\alpha$, and Pa\,$\alpha$ kinematics within $\approx2$ effective radii or $\approx 7$\,kpc (depending on the exact model adopted;  well-matched within error to the definition of enclosed mass in the simulations), as described in detail in \cite{Hodge2012, Bik2024, Ubler2024}, based on data from the Jansky Very Large Array, JWST/NIRSpec IFU, and JWST/MIRI MRS, respectively.

We re-estimate the total stellar mass of GN20 from spectral energy distribution (SED) modelling of the extensive UV--FIR photometry, now including the new JWST NIRCam and MIRI observations.  We use \textsc{magphys} \citep{DaCunha2008,DaCunha2015,Battisti2020} that incorporates stellar templates from \cite{Bruzual2003} with an  initial mass function from \cite{Chabrier2003} and a dust model from \cite{Charlot2000} with approximate energy balance between the light absorbed and re-emitted by dust at long wavelengths.  The HST and JWST photometry was measured directly from the maps using a circular aperture with a radius of $r = 1.4''$, while the (unresolved) far-infrared photometry was adopted from \cite{Crespo-Gomez2024} and \cite{Boogaard2026}, accounting for up to 0.2\,dex of systemic error in the flux calibration.  The stellar mass estimate is $\log(M_* / {\rm M}_{\odot}) = 11.1 \pm 0.25$, in excellent agreement with the earlier estimate from \cite{Tan2014} incorporating the unresolved Spitzer IRAC and MIPS photometry, which is not surprising given that GN20 is relatively isolated.  Given the AGN signatures, we also estimate the mass using an adopted version of the \textsc{magphys+agn} model \citep{Chang2017} finding $\log(M_* / {\rm M}_{\odot}) = 11.0 \pm 0.25$ with a moderate global AGN contribution of $\leq20$\,\%.  For comparison, we also perform an independent estimate of the stellar mass using \textsc{cigale} with a \cite{Calzetti2000} dust law, delayed-$\tau$ SFH, and including an AGN component using \cite{Fritz2006} and nebular emission templates from \cite{Inoue2011}, finding a value of $\log(M_* / {\rm M}_{\odot}) = 11.06 \pm 0.2$ which
is slightly higher than the estimate from \cite{Crespo-Gomez2024} who estimated a total stellar mass of $M_{\rm star} = 10.9 \pm 0.22 \times 10^{10}$\,M$_{\odot}$ from the same code.  It is well understood that different methods for modelling SEDs and detailed processing of the photometry can give slight systematic differences \citep[e.g.,][]{Pacifici2023}.  For the purpose of this work, we adopt a stellar mass of $1.0 \pm 0.5 \times 10^{11}$\,M$_{\odot}$, based on the \textsc{magphys+agn} model.  The best-fit integrated SED model from \textsc{magphys} is shown as the black curve in \autoref{fig:magphys_best-fit}, along with additional SED modelling of just the core and bar region using more limited photometry, which is described below.  The integrated physical properties derived from all SED fits are listed in the first column of \autoref{tab:sed_aperture}.

The total molecular gas mass of GN20 has been recently estimated to be
$M_{\rm mol} = 2.9 \pm 0.4 \times 10^{11}$\,M$_{\odot}$, based on
novel radiative transfer modeling of all the available multi-$J$ $^{12}$CO lines and multi-frequency dust continuum emission \citep{Boogaard2026}.  This is consistent with a
straightforward mass estimate from the luminosity of the single
$^{12}$CO $J_{\rm up}=1\rightarrow0$ transition for an
$\alpha_{\rm CO}\approx 1.8$ \citep[cf.][see the extensive discussion in \citealt{Boogaard2026}]{Carilli2010, Hodge2012}. We note that using the commonly-adopted, but poorly motivated, value of $\alpha_{\rm CO} =0.8$ would still leave the galaxy very gas rich and not significantly affect our conclusions.

\section{Theory and simulations of bar formation in early gas-rich disks}\label{sec:theory}
\begin{figure*}
  \centering
  \includegraphics[width=\textwidth]{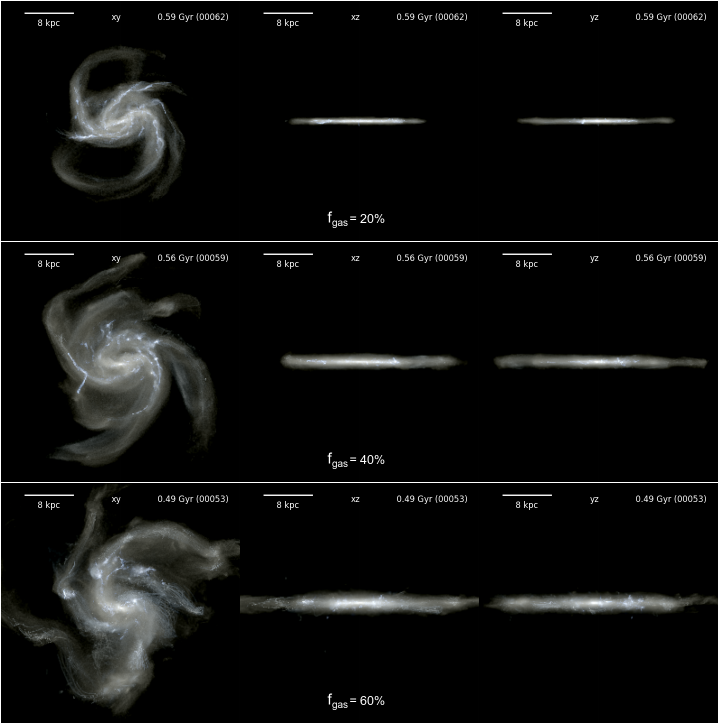}
  \caption{Morphology of GN20-like simulated barred galaxies.
    Multiple projections of stellar bars forming in simulated gas-rich
    disk galaxies at different gas fractions \citep{Tepper-Garcia2024,
      Bland-Hawthorn2024, Bland-Hawthorn2025}.\label{fig:sims}}
\end{figure*}

\begin{figure}
  \centering
  \includegraphics[width=\columnwidth]{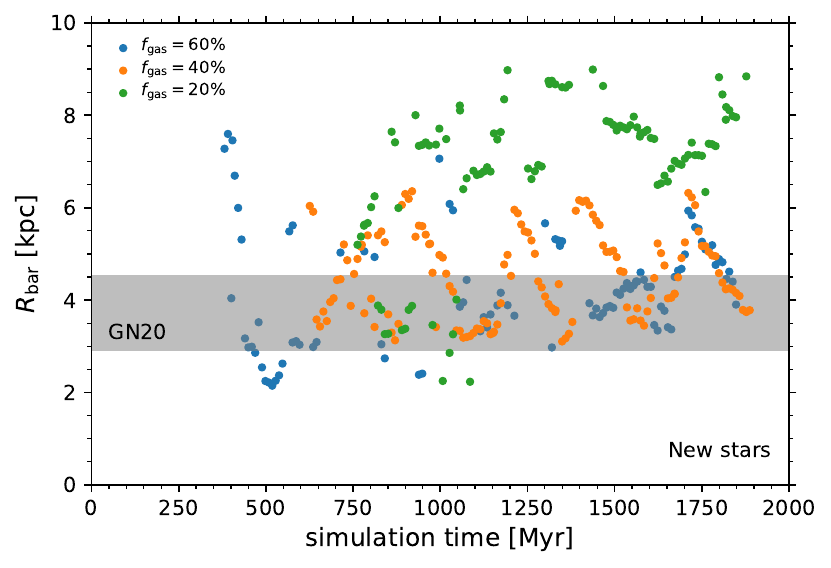}
  \caption{Stellar bar length evolution in cosmological simulations.
    Bar radius as function of simulation time, as derived via the
    Dehnen/Fourier method (see \autoref{sec:morph}).  The GN20 bar radius
    ($R_{\rm bar/\phi_2}$, derived consistently) matches most closely
    with the simulations with high gas fractions ($f_{\rm gas} = 60$\%
    and 40\%). \label{fig:theory_ext}}
\end{figure}

\subsection{Simulations of turbulent gas-rich disks at high redshift}
In a series of earlier papers, a new high-resolution (parsec-scale) simulation framework, \textsc{Nexus}, was introduced specifically to study the dynamical evolution of turbulent discs in early galaxies \citep{Tepper-Garcia2024}. These systems exhibit both intermittent and long-lived behaviour, including disc–halo interactions, the formation of bars and spiral arms, and the emergence of central bulges \citep{Bland-Hawthorn2024}. They demonstrated for the first time that bar-like phenomena can arise across the full range of gas fractions, even in fully gas-dominated turbulent discs.

Within the \textsc{Nexus} framework, there are two key parameters that may explain the existence of early massive stellar bars:
(i) The disc mass fraction, $f_{\rm disc}$, which determines whether the disc baryons dominate the underlying dark matter halo,
\begin{eqnarray}
\label{e:fd}
    f_{\rm disc} = \left(\frac{V_{\rm c, disc}(R_s)}{V_{\rm c, tot}(R_s)}\right)_{R_s=2.2 R_{\rm disc}}^2 \;.
\end{eqnarray}
Here, $V_c(R)$ is the circular velocity at a radius $R$, $R_{\rm disc}$ is the exponential disc scale length, and $R_s=2.2 R_{\rm disc}$ is the traditional scale length adopted in studies of discs; (ii) The gas mass fraction within the same radial scale:
\begin{eqnarray}
\label{e:fdd}
    f_{\rm gas} =
    \left(\frac{M_{\rm disc, gas}(R_s)}{M_{\rm disc}(R_s)}\right)_{R_s=2.2 R_{\rm disc}}
\end{eqnarray}
where $M_{\rm disc}$ is the total disc mass and $M_{\rm disc,gas}$ is the gas contribution.

All computations were carried out with the \textsc{Ramses} N-body/hydrodynamics code at benchmark ($N\sim 10^7$) including star formation and metal production; the number of effective gas ``cells'' is a factor of 10 higher in both cases. In defining a massive progenitor at $z\approx 4$, we adopted a model with three key components: a live dark matter halo, a massive stellar/gaseous disc, and a hot coronal halo.
For our massive progenitor, we adopted halo parameters of $R_{\rm vir}\approx 40$ kpc and $\log M_{\rm vir}/{\rm M}_\odot \approx 11$.

We ran simulations for 2~Gyr at $f_{\rm disc}=70\%$ with respective gas fractions of $f_{\rm gas} = (0,20,40,60) \%$,  with and without halo cooling \citep{Tepper-Garcia2024}. The cooling halo models allow for additional gas accretion onto the disc, but because $f_{\rm disc}$ is so extreme, the additional baryons made little difference compared to what was observed in our earlier $f_{\rm disc}=50\%$ models \citep[cf.][]{Bland-Hawthorn2024}. For the halo accretion models to be substantially different from the non-accreting models, we would have needed a coronal gas mass that exceeded the cosmic baryon fraction ($f_{\rm cos}\approx 16\%$).

Representative snapshots are shown for the three non-accreting cases in \autoref{fig:sims}.  We note that GN20 (just as the systems from \citealt{Huang2025, Amvrosiadis2025}) is still somewhat more massive galaxy in terms of its disk and virial (halo) mass than the simulated galaxy.  This does not affect any of our conclusions, as the driving parameters are $f_{\rm disc}$ and $f_{\rm gas}$ and the key physics is intact, but we do caution the precise quantities (such as bar length) depend on the modelled galaxy properties (including mass and size).

Synthetic galaxy images (\autoref{fig:theory} and \ref{fig:sims}) were generated using the package {\small Pynbody} \citep{Pontzen2013}, which calculates the luminosities of star particles based on stellar population synthesis models. Each particle in the simulation represents a simple stellar population (SSP), corresponding to a coeval population of stars with a single age and metallicity.  The intrinsic magnitudes in photometric bands are obtained by interpolating precomputed SSP tables, and scaled according to the stellar mass of each particle, such that the luminosity of a particle is directly proportional to its mass. This ensures that the resulting photometry captures the dependence of colour and brightness on both age and chemical composition of the stellar population, with younger, more metal-poor populations appearing bluer and more luminous, and older, metal-rich populations appearing redder and fainter.  To construct false-colour images, the particle luminosities are projected onto a two-dimensional grid, producing surface brightness maps in multiple photometric bands. Magnitudes are converted to linear fluxes, which are then mapped to RGB channels to create composite images, allowing the visualization to highlight specific stellar population properties or spectral features similar to real observations.  Given the highly complex dust attenuation properties in GN20, we purposefully do not include the effect of dust attenuation in the synthetic images, which thus provide a more direct comparison to the redder JWST filters that are less attenuated.  The RGB channels shown in \autoref{fig:theory} and \ref{fig:sims} correspond to the Johnson I, V and B band respectively, with the reddest band corresponding to approximately 4\,$\mu$m observed-frame (close to NIRCam F444W, see \autoref{fig:rgb}).

We use Dehnen's method \citep{Dehnen2023} to extract the bar parameters presented in \autoref{fig:theory} and \autoref{fig:theory_ext}. The disc's surface density $\Sigma$ is decomposed into azimuthal Fourier modes:
\begin{equation}
A_m(R) = \frac{1}{M(R)} \sum_j m_j \, e^{i m \phi_j},
\end{equation}
where $R$ is the cylindrical radius, $\phi_j$ is the azimuthal angle of particle $j$, $m_j$ is the particle mass,
and $M(R)$ is the total mass in the radial bin.
The radial bar strength is defined as the normalized $m=2$ amplitude: $|A_2(R)|$ where by convention we drop the modulus symbols and normalize using $A_2(R)/A_0(R)$. The phase of the $m=2$ mode is $\phi_2(R) = \frac{1}{2} \arg\left(A_2(R)\right)$. Within the bar region, $\phi_2(R)$ is approximately constant, defining the bar orientation. The bar radius $R_{\mathrm{bar}}$ is defined as the largest radius for which the $m=2$ phase remains coherent, i.e.
\begin{equation}
\left| \phi_2(R) - \phi_{\mathrm{bar}} \right| < \Delta \phi,
\end{equation}
where $\phi_{\mathrm{bar}}$ is the mean inner bar phase and $\Delta \phi \approx 10^\circ$. The bar pattern speed is obtained from the time evolution of the bar phase:
$\Omega_{\mathrm{bar}} = d\phi_\mathrm{bar}/dt$.

\subsection{Theory and discussion}
High-redshift stellar bars ($z\gtrsim 4$) are a remarkable phenomenon because they indicate that the processes capable of assembling and stabilizing discs were already operating efficiently within the first billion years of cosmic history. GN20 hosts a dynamically cold, massive ($\approx 5\times 10^{11}$ M$_\odot$) disc that is dominated by baryons largely in the form of gas ($f_{\rm gas}\gtrsim 50\%$). From the gas kinematics, the baryons dominate over dark matter across the inner disc with $f_{\rm disc}\approx 70\%$.

Given the observational facts outlined above, it is remarkable that such a massive bar has formed at all. In a disc-dominated system, bar formation is expected to occur readily \citep{fuj18a}; however, the resulting bar is typically so strong that it becomes vertically unstable and weakens or destroys itself through buckling \citep{raha91}. Additional challenges remain. The growth of a long stellar bar is intrinsically slow, proceeding at only $\sim 0.6$ kpc Gyr$^{-1}$ \citep{fuj18a}, and the presence of gas is generally thought to further delay bar formation \citep{Athanassoula2013}. Our new results demonstrate that all three of these obstacles can be overcome by a {\it single} ingredient directly implicated by the observations: {\it the presence of highly turbulent gas across the inner disc at high gas fraction.}

In \autoref{fig:theory}, we present the Fourier analysis of all four \textsc{Nexus} simulations. Remarkably, for the first time, we see a clear dependence of the emergence of the bar as a function of cosmic time over a wide range in gas fraction. Broadly speaking, the functional dependence is similar across all gas fractions, differing only in onset time and main peak Fourier amplitude $A_2/A_0$. In a separate paper, we present a simple analytic model that explains all three properties, i.e. near constant growth rate, decreasing $A_2/A_0$ peak amplitude, and onset time with $f_{\rm gas}$ (Bland-Hawthorn et al. 2026, MNRAS, submitted).

\cite{Bland-Hawthorn2023} showed that the normalized amplitude of the simulated bar, $A^\prime_2 = A_2(t-\tau_0)/A_0$, is well approximated by an exponential dependence on time $t$, in agreement with theoretical expectations for a positive feedback loop operating on instabilities seeded by a noisy, self-gravitating distribution. We can include the contribution of a stochastic process, i.e.
\begin{eqnarray}
    \frac{dA^\prime_2}{dt} = \gamma_\star A^\prime_2(t) + \eta_{\rm gas}(t)
    \label{e:lang}
\end{eqnarray}
where $\gamma_\star=1/\tau_{\rm exp}$ is the rate of growth and $\eta_{\rm gas}$ describes the source of noise fluctuations in the gravitational potential $\Phi$. For a stochastic process,
\begin{eqnarray}
    \langle\eta(t)\rangle &=& 0\\
     \langle\eta(t)\eta(t^\prime)\rangle &=& {\cal D}\: \mathbf{\delta}(t-t^\prime)
\end{eqnarray}
where ${\cal D}$ is the diffusion coefficient and $\mathbf{\delta}$ is the Dirac delta function. An elegant aspect of the model is that the outcome depends specifically on the measured surface density fluctuations $\delta\Sigma$ in the \textsc{Nexus} simulations \citep{zhang25}, i.e.
\begin{eqnarray}
     \eta_{\rm gas} = -\nabla\: \frac{\delta\Phi_{\rm gas}}{\Phi_{\rm tot}} \approx - 2\pi G f_{\rm gas}\: \frac{\delta\Sigma_{\rm tot}}{\Sigma_{\rm tot}}
     \label{e:force}
\end{eqnarray}
where the fluctuations are averaged over an orbit. This model leads to the three desired outcomes mentioned above for the onset time, growth rate, and $m=2$ Fourier amplitude.

\section{Triggering, timescales and properties of the stellar bar}\label{sec:timescale}
We investigate whether past- or on-going interactions of GN20 with companion galaxies in the close vicinity could have have triggered the bar instability, in addition to the possibility of baryon sloshing.  The main companion galaxy GN20.b \citep[][see \autoref{fig:rgb}]{Ubler2024}  is at a projected, and thus minimal, distance of approximately 15\,kpc and a relative velocity of 750\,km\,s$^{-1}$, which is too fast and far away to exert significant torques on the GN20 disk.  In addition, smaller (blue) clumps are detected to the south-east and south-west of GN20, as well as north-west of the spiral arm, that could correspond either to star-forming clumps near the disks, or nearby companions.  In the latter case, these companions are at closer proximity and closer to the systemic velocity of GN20.  In addition, we cannot rule out that GN20 is in the post-coalescence phase of a major-merger, which are known to trigger periods of intense star formation and possibly also the onset of a bar.  Both merger activity and baryon-sloshing are consistent with the asymmetric structure of the GN20 disk, including the $m=1$ spiral arm and offset-nucleus \citep{Colina2023}.  In this context, we note that the starburst at bar-disk interface in GN20, also often seen in local galaxies \citep[e.g.,][]{Beuther2017}, is also asymmetric, occurring only in the south, which is more uncommon.  Such asymmetry can be explained from baryon sloshing and has also been associated with off-centered nuclei (e.g., \citealt{Sanchez-Martin2023}).  The stellar bar is relatively long compared to the total size of the disk.  This long bar length appears consistent with predictions from \cite{Fragkoudi2025} that at high-$z$ bars form saturated in length, while (lower-$z$) bars formed through secular evolution grow in time.

To provide an approximate analysis of the bar and core properties and the timescale of the bar build-up, in addition to the integrated SED modelling, we perform SED modelling of the core and bar regions (North and South) separately, using the same codes and methodologies as described above, but using only the, more restricted, resolved photometric data.  We perform aperture photometry on the regions shown in \autoref{fig:magphys_best-fit}, and the best-fit models for these three regions are displayed as the red, green, and cyan curves, along with the integrated SED model in black.  The mass-weighted ages derived for the bar are of order 300 Myr (albeit with significant uncertainty), which is consistent with the estimates from the simulations.  The stellar mass of the core and bar regions are substantial (of order a few times $10^{10}$\,M$_{\odot}$), with the best-fit models showing approximately 4--5 magnitudes of visual attenuation.  We note that the mass estimates of the core and bar region together approach the estimated integrated mass estimate of the system.  To further investigate this, we also perform spatially resolved SED modelling on a pixel-by-pixel basis across the entire galaxy \citep[see][and C.L.~Liao et al. (subm.) for a description of the methodology]{Li2024}, using the same data matched to the filter with the coarsest angular resolution, MIRI/F560W ($\sim0.21''$), but without the AGN models.  This results in a broadly consistent mass estimate for the core and bar regions, compared to the aperture method, but a larger stellar mass for the entire system, rivalling the mass in the molecular gas component.  If real, this discrepancy could be due to a variation of the `outshining' effect, where the less-obscured outskirts of GN20 outshine the dust-obscured central regions, thereby biasing the fit towards lower extinction and thus lower mass (as reflected in the lower $A_V$ values).  However, the high inferred stellar mass is at tension with (exceeds) the total dynamical mass of the system (albeit with large uncertainties), and we caution these fits are obtained without the AGN models (which would likely bring down the mass) and with more limited photometry, in particular lacking the mid-infrared and the detailed shape of the far-infrared dust SED.  Future work on the resolved SED modelling is required to shed more light on this.  In any case, while a resulting higher stellar mass estimate would bring down the gas fraction, it remains consistent with GN20 being a gas-rich system, and would not affect the main conclusions of the paper.

\bibliography{biblio}
\bibliographystyle{aasjournalv7}

\end{document}